\documentstyle[prl,aps,multicol]{revtex}

\def\re{{\rm Re}}
\def\im{{\rm Im}}

\title{Minigap in a long disordered SNS junction: analytical results}

\author{Dmitri~A.~Ivanov, Raphael von Roten, and Gianni Blatter}

\address{Institut f\"ur Theoretische Physik,
ETH-H\"onggerberg, CH-8093 Z\"urich, Switzerland}

\date{April 2, 2002}

\begin{document}

\maketitle

\begin{abstract}

We review and refine analytical results on the density of states
in a long disordered
super\-conductor--normal-metal--super\-conductor junction with
transparent interfaces. Our analysis includes the behavior of the
minigap near phase differences zero and $\pi$ across the junction,
as well as the density of states at energies much larger than the
minigap but much smaller than the superconducting gap.

\end{abstract}

\begin{multicols}{2}

A superconductor in contact with a normal metal induces pairing
correlations in the metal, a phenomenon known as the proximity
effect. One of the most remarkable consequences of such induced
correlations is the appearance of a gap, usually referred to as
the ``minigap'', in the electronic excitation spectrum of the
normal metal~\cite{McMillan-Belzig}. A very common setup
exhibiting a minigap is the
super\-conductor--normal-metal--super\-conductor (SNS) junction
made of two superconducting leads connected via a disordered
normal layer. The gap in this junction is determined by the
diffusion time across the normal layer and is sensitive to the
phase difference across the junction, reaching a maximum at zero
phase difference and vanishing at the phase difference
$\pi$~\cite{Charlat}.

The appearance of a minigap and its phase dependence is well
understood in the quasiclassical description. Provided the
scattering length is much larger than the Fermi wave length but
much smaller than the junction dimensions, the motion of the
electrons in the normal layer is diffusive and the proximity
effect may be described by the Usadel equations~\cite{Usadel}.
These equations are nonlinear, which complicates their analytical
treatment except for several simple limits. One of the cases most
accessible to an analytical treatment is the limit of a long
disordered SNS junction (with the minigap energy scale much
smaller than the superconducting gap) with transparent
normal-metal--superconductor interfaces. The spectral properties
of such a junction have been previously studied in
Refs.~\onlinecite{Charlat,Charlat-thesis} and we find it possible
to further improve on those results. In this note we revisit this
problem, refining some of the existing results and replacing
numerical answers with analytical ones. We design this note as a
quick reference on the structure of the minigap in a long
disordered SNS junction which may be useful in view of renewed
interest in such systems in connection with problems related to
$\pi$-junctions~\cite{Bulaevskii-Buzdin,Ryazanov1} and to
mesoscopic fluctuations~\cite{Altland,Ostrovsky}. As a byproduct,
we derive two useful identities for solutions to Usadel equations
which simplify our analytical calculations.

Assuming a quasi-one-dimensional geometry of the contact, the
proximity effect in the normal layer may be described via 
the Usadel equations (in our paper we
conform to the definitions of Ref.~\onlinecite{Charlat})
\begin{eqnarray}
   && {1\over2} \partial_x^2 \theta + i\varepsilon\, \sin\theta
   -{1\over4} (\partial_x \chi)^2 \sin 2\theta
   = 0 \, , \nonumber \\
   && \partial_x\left( \partial_x\chi \, \sin^2 \theta \right)
   =0 \, ,
   \label{Usadel-eqns}
\end{eqnarray}
where $\theta(x,\varepsilon)$ and $\chi(x,\varepsilon)$ are the
variables parameterizing the zero angular momentum component of
the Green's functions, $g = \cos\theta$ and $f = \sin\theta\,
\exp(i\chi)$; here, $\chi$ is the phase of the superconducting
correlations, and the local density of states $\rho(x,\varepsilon)$
(in units of the normal electron density in the bulk) is given by
\begin{equation}
   \rho(x,\varepsilon) = \re\, \cos\theta\, .
   \label{DOS-definition}
\end{equation}
Note that we measure lengths in units of the junction width $L=1$
and the unit of energy $\varepsilon$ is the Thouless energy
$E_c=D/L^2$, with $D$ the diffusion constant in the normal metal.
We further assume that the Thouless energy $E_c$, as well as all
other energy scales in the problem, are much smaller than the
superconducting gap $\Delta$. In this limit, the energy scale
$\Delta$ is (to leading order) excluded from the Usadel equations.

Within the superconducting leads, $\theta=\pi/2$ (at $\varepsilon
\ll \Delta$) and $\chi$ equals the phase of the superconducting
order parameter. Inside the junction, both $\theta$ and $\chi$
turn into complex functions. For simplicity, we choose ideally
transparent interfaces and assume the normal metal to be much more
disordered than the superconductor; in this case the boundary
conditions become ``rigid''~\cite{Altland,Lukichev},
\begin{eqnarray}
   && \theta(x=0)=\theta(x=1)={\pi\over2}\, , \nonumber \\
   && \chi(x=0)=0\, ; \qquad \chi(x=1)=\chi_0\, .
   \label{boundary-conditions}
\end{eqnarray}

The equations (\ref{Usadel-eqns})--(\ref{boundary-conditions})
form a closed set determining the density of states (with $\chi_0$
and $\varepsilon$ as input parameters). They have been analyzed in
Refs.~\onlinecite{Charlat,Charlat-thesis} and in this note we
extend their results; we list them first and sketch their derivation
afterwards.

\bigskip

{\bf (i)} The complex integral ``density of states'' defined as
\begin{equation}
   \hat\rho(\varepsilon)= \int_0^1 dx\, \cos\theta(x,\varepsilon)
\end{equation}
obeys the relations
\begin{eqnarray}
   \hat\rho+2\varepsilon {\partial\hat\rho\over\partial\varepsilon}
   &=& i{\partial B\over\partial\varepsilon}\, ,
   \label{integral-B} \\
   2{\partial\hat\rho\over\partial\chi_0}
   &=& -i{\partial C\over\partial\varepsilon}\, ,
   \label{integral-C}
\end{eqnarray}
where $B(\varepsilon,\chi_0)$ and $C(\varepsilon,\chi_0)$ are
integrals of (\ref{Usadel-eqns}),
\begin{eqnarray}
   B &=& {1\over4} (\partial_x\theta)^2 - i\varepsilon \cos\theta
   +{C^2 \over 4\sin^2\theta}\, ,
   \label{B-def}\\
   C &=& (\partial_x \chi)\, \sin^2\theta\, .
   \label{C-def}
\end{eqnarray}

{\bf (ii)} Near $\chi_0 \to 0$ and $\chi_0 \to\pi$, 
the phase dependence of the minigap
$E_g(\chi_0)$ (in units of $E_c$) involves the leading terms,
\begin{eqnarray}
   E_g(\chi_0) &=& C_2(1-C_1 \chi_0^2)\, , \qquad \chi_0 \ll \pi\, ,
   \label{Eg-chi-0}\\
   E_g(\chi_0) &=& C_3 (\pi-\chi_0)\, ,    \qquad \pi-\chi_0 \ll \pi\, .
   \label{Eg-chi-pi}
\end{eqnarray}
The value of $C_2$ has been derived in \onlinecite{Charlat},
\begin{equation}
   C_2=\left[\max_{\vartheta_0} \int_0^{\vartheta_0} {d\vartheta\over
   (\sinh\vartheta_0-\sinh\vartheta)^{1/2}}\right]^2 \approx 3.122\, ,
   \label{C2}
\end{equation}
with the maximum attained at $\hat{\vartheta}_0 \approx 1.421$.
The value of $C_1$ reported in \onlinecite{Charlat} is incorrect
by an order of magnitude and we find the correct value
\begin{equation}
   C_1 = {\int_0^{\hat{\vartheta}_0}\!\!
   {d\vartheta\, (\sinh\hat{\vartheta}_0 + \sinh\vartheta)
   \over
   (\sinh\hat{\vartheta}_0 - \sinh\vartheta)^{1/2} \cosh^2\vartheta\,
   \cosh^2\hat{\vartheta}_0}
   \over
   4\sqrt{C_2} \left[
   \int_0^{\hat{\vartheta}_0}\!\! {d\vartheta \over
   (\sinh\hat{\vartheta}_0 - \sinh\vartheta)^{1/2} \cosh^2\vartheta}
   \right]^2}
   \approx 0.0921\, .
   \label{C1}
\end{equation}
For $C_3$ we find the analytic result close to the numerical
value reported in \onlinecite{Charlat},
\begin{equation}
   C_3={\pi^2\over4}\approx 2.467\, .
   \label{C3}
\end{equation}

{\bf (iii)} The integral density of states $\rho(\varepsilon)=\re\,
\hat\rho(\varepsilon)$ has a square-root singularity at
$\varepsilon=E_g$~\cite{Charlat},
\begin{equation}
   \rho(\varepsilon) \sim C_4(\chi_0) \left({\varepsilon-E_g
   \over E_g} \right)^{1/2}\, ,
   \label{gap-edge}
\end{equation}
with the coefficient $C_4$ diverging at $\chi_0 \to \pi$ as
\begin{equation}
   C_4(\chi_0) = \alpha (\pi-\chi_0)^{-2/3}\, .
\end{equation}
This asymptotic form has been found numerically in Refs.\
\onlinecite{Charlat,Charlat-thesis} and we confirm it here
analytically, together with the value for $\alpha$,
\begin{equation}
   \alpha={1\over \pi\sqrt6}
   \left({1\over 2\pi} - {3\pi\over 64}\right)^{-2/3}
   \!\! \approx 2.494\, .
   \label{alpha-coefficient}
\end{equation}

{\bf (iv)} At energies $\varepsilon$ much higher than $E_c$ but much
lower than the superconducting gap, the leading corrections to the
(integral) density of states are
\begin{equation}
   \rho(\varepsilon, \chi_0) \approx 1 - \frac{C_5}{\sqrt{\varepsilon}}
   + e^{-\sqrt\varepsilon} C_6(\varepsilon) \cos(\chi_0)\, ,
   \label{high-energy-rho}
\end{equation}
where
\begin{equation}
   C_5 = 2- \sqrt2 \approx 0.586\, ,
   \label{C5}
\end{equation}
\begin{equation}
   \label{C6}
   C_6(\varepsilon) 
= 16 \tan^2{\pi\over 8}
   \biggl[-\cos\sqrt\varepsilon
   +\frac{\cos\sqrt\varepsilon
   -\sin\sqrt\varepsilon}{2\sqrt\varepsilon}\biggr]  \; .
\nonumber
\end{equation}

\bigskip

The verification of  {\bf (i)} is straightforward: We integrate the Usadel
equations (\ref{Usadel-eqns}) once and denote the
coordinate-independent integrals by $B$ and $C$ as in
(\ref{B-def}) and (\ref{C-def}). The relations (\ref{integral-B})
and (\ref{integral-C}) are then obtained via integration by parts
in $x$ and repeated use of the Usadel equations, see the Appendix
for details. Relation (\ref{integral-B}) expresses the
conservation of the total number of states as a function of
$\chi_0$; indeed, (\ref{integral-B}) implies that
$\hat\rho(\varepsilon)$ is a total derivative in energy of the
expression $(2\varepsilon\hat\rho-iB)$ which is independent of
$\chi_0$ at large energies [this follows from our discussion of
the result $iv)$ below]; therefore, the integral of
$\hat\rho(\varepsilon)$ over energies is independent of $\chi_0$.
The identity (\ref{integral-C}) follows from the fact that the
supercurrent (which equals $\im\, C$ in appropriate
units~\cite{Usadel}) can be expressed as the derivative of the free
energy with respect to $\chi_0$; we verify (\ref{C-def}) directly
on the level of Usadel equation in the Appendix.

To obtain the results of paragraph {\bf (ii)} we follow the usual
procedure in integrating the Usadel equations: below the gap,
$\theta(x)$ takes the form $\theta=\pi/2 + i\vartheta$, where
$\vartheta(x)$ is real and varies from $0$ at the N--S interface
to its maximal value $\vartheta_0$ in the middle of the normal
layer. From (\ref{B-def}) and (\ref{C-def}) we obtain the differential
equations
\begin{equation}
   \partial_x \vartheta
   =\sqrt{f(\vartheta_0)-f(\vartheta)}/\sqrt{\varepsilon}\, ,
   \quad
   \partial_x\chi
   =C \cosh^{-2} \vartheta\, ,
   \label{de}
\end{equation}
where $f(\vartheta)=\sinh(\vartheta)-(C^2/4\varepsilon) \cosh^{-2}
\vartheta$. Their integration over half the junction provides us
with the solutions of the Usadel equations in the form
\begin{eqnarray}
   \sqrt\varepsilon
   &=& \int_0^{\vartheta_0}
   {d\vartheta \over [f(\vartheta_0)-f(\vartheta)]^{1/2}}\, ,
   \nonumber \\
   \chi_0
   &=& {C\over\sqrt\varepsilon} \int_0^{\vartheta_0}
   {d\vartheta \over \cosh^2\vartheta \,
   [f(\vartheta_0) - f(\vartheta)]^{1/2}}\, ,
   \label{new-equations}
\end{eqnarray}
expressing $\varepsilon$ and $\chi_0$ in terms of the new
parameters $C/\sqrt{\varepsilon}$ and $\vartheta_0$. The minigap
$E_g(\chi_0)$ is defined as the maximal energy $\varepsilon$
compatible with this solution (with a real $\vartheta(x)$).
Equation (\ref{C1}) is now easily obtained by expanding in $C$
which is the small parameter near $\chi_0 = 0$, while the result
(\ref{C3}) is obtained from an expansion in $\varepsilon$, the
small parameter close to $\chi_0 = \pi$ where the minigap vanishes
(see also the derivation below).

A more accurate expansion in small $\varepsilon$ is necessary to
obtain the results of paragraph {\bf (iii)}. The density of states near
the gap edge (\ref{gap-edge}) is derived with the help of relation
(\ref{integral-B}). The key idea of the calculation is that most
quantities are regular functions of the new parameters $\vartheta_0$
and $\varepsilon/C^2$ [except for the point $\chi_0=\pi$ where the
minigap vanishes]. The square-root singularity (\ref{gap-edge})
appears when inverting regular functions at the extremal point.
Remarkably, in this way we can compute the density of states at
the gap edge using the solutions to the Usadel equations {\it
below} the gap and analytically continuing them to energies above
the gap.

Close to $\chi_0 = \pi$ the energy $\varepsilon$ is small and we
can expand (\ref{new-equations}) in the small parameter
$\delta=\varepsilon/C^2$,
\begin{equation}
   \varepsilon = \left( \pi^2 \cosh^2 \vartheta_0 \right) \delta
   -\pi\left({1\over 2}-{\pi\over 8}\right)
   e^{5\vartheta_0} \delta^2 + \dots\, ,
   \label{epsilon-x}
\end{equation}
\begin{equation}
   \chi_0=\pi - {4\cosh^3\vartheta_0\over \sinh\vartheta_0} \delta
   + \left({3\pi\over 16}-{1\over 2} \right)
   e^{5\vartheta_0} \delta^2 + \dots.
   \label{chi-x}
\end{equation}
We will show below that the solutions at the gap edge near
$\chi_0=\pi$ involve a large parameter $\vartheta_0 \gg 1$, and
hence we may keep only the leading terms in $e^{\vartheta_0}$ in
the coefficients of $\delta^2$. Next we invert (\ref{chi-x}) to
find
\begin{equation}
   \delta \approx
   {\sinh \vartheta_0 \over 4 \cosh^3 \vartheta_0} (\pi-\chi_0)
   +\biggl[{3\pi\over 16} -{1\over 2} \biggr]
   e^{-\vartheta_0} (\pi-\chi_0)^2
   \label{x-chi}
\end{equation}
and substitute the result into (\ref{epsilon-x}) to express the
energy $\varepsilon$ as a power series in $(\pi-\chi_0)$ and as a
function of $\vartheta_0$,
\begin{equation}
   \varepsilon \approx
   {\pi^2\over 4} (\pi-\chi_0)\tanh\vartheta_0
   -\biggl[{\pi\over 2}\!-\!{3\pi^3\over 64}\biggr] e^{\vartheta_0}
   (\pi-\chi_0)^2\, .
   \label{epsilon-theta-chi}
\end{equation}
The gap edge is given by maximizing $\varepsilon$ in
(\ref{epsilon-theta-chi}) as a function of $\vartheta_0$; the
corresponding value $\bar{\vartheta}_0$ maximizing $\varepsilon$
is
\begin{equation}
   \bar{\vartheta}_0={1\over 3} \ln{\chi_c\over\pi-\chi_0}, \quad
   \chi_c=\left({1\over 2\pi} - {3\pi\over 64}\right)^{-1}\!\!
   \approx 84.08\, ;
   \label{theta-pound}
\end{equation}
we see that $\bar{\vartheta}_0$ is indeed large at the gap edge
(for small $\pi-\chi_0$), albeit only logarithmically and with a
small pre\-factor $1/3$. Upon substitution into
(\ref{epsilon-theta-chi}), this result also gives us the next-order
correction to (\ref{Eg-chi-pi}),
\begin{equation}
   E_g(\chi_0)
   = {\pi^2\over 4} (\pi-\chi_0)\left[1-6\left({\pi-\chi_0 \over
   \chi_c}\right)^{2/3}\right]\, .
\end{equation}

We are now prepared to derive the density-of-states singularity at
the gap edge (\ref{gap-edge})--(\ref{alpha-coefficient}).
The maximum of the function $\varepsilon(\vartheta_0)$ at
$\bar{\vartheta}_0$ is expressed in the relation
$\varepsilon(\vartheta_0) \approx E_g
- |\partial_{\vartheta_0}^2 \varepsilon|_{\bar{\vartheta}_0}\,
[\bar{\vartheta}_0-\vartheta_0]^2/2$; its inversion produces a
square-root singularity at the gap edge in the function
$\vartheta_0(\varepsilon)$,
\begin{equation}
   \vartheta_0(\varepsilon)=\bar{\vartheta_0} \pm i
   \sqrt{\frac{2(\varepsilon-E_g)}
   {|\partial_{\vartheta_0}^2 \varepsilon|_{\bar{\vartheta}_0}}}\, ,
\end{equation}
and hence $\vartheta_0$ develops an imaginary part at energies
$\varepsilon > E_g$. The singularity in $\vartheta_0(\varepsilon)$
translates into a square-root singularity in $B(\varepsilon)$;
evaluating (\ref{B-def}) in the junction middle and expressing $C$
with the help of (\ref{x-chi}), we obtain
\begin{eqnarray}
   B&=&\varepsilon\left[{\coth\vartheta_0 \over \pi-\chi_0}
   -{3\pi\over 16} e^{\vartheta_0}\right]
   \approx
   \re B\pm i\varepsilon\sqrt{\frac{2(\varepsilon-E_g)}
   {|\partial_{\vartheta_0}^2 \varepsilon|_{\bar{\vartheta}_0}}}
   \nonumber \\
   &\phantom{.}& \quad \times
   \bigl[(\pi-\chi_0)^{-1}\sinh^{-2}\bar{\vartheta}_0
   + 3\pi e^{\bar{\vartheta}_0}/16\bigr]\; .
   \label{B-singularity}
\end{eqnarray}
This singularity in $B$ further translates, via (\ref{integral-B}), into a
square-root singularity in the density of states
$\rho(\varepsilon)$. Evaluating the coefficient in
(\ref{B-singularity}) with the use of (\ref{epsilon-theta-chi})
for $\varepsilon(\vartheta_0)$ and of (\ref{theta-pound}) for
$\bar{\vartheta}_0$, we arrive at the final results
(\ref{gap-edge})--(\ref{alpha-coefficient}). The coefficient
(\ref{alpha-coefficient}) agrees with the numerical findings in
\onlinecite{Charlat}.

Another regime where the density of states is amenable to a simple
analytic solution is at energies much larger than $E_c$ but much
smaller than $\Delta$. In this limit, the coupling between the
superconducting leads is weak, which allows us to derive the
results of paragraph {\bf (iv)}. In equation (\ref{high-energy-rho}),
the term proportional to $\varepsilon^{-1/2}$ is due to the
suppression of the density of states near the interfaces, and the
exponentially small term proportional to $\cos\chi_0$ results from
the Josephson coupling.

At $\varepsilon\gg 1$, the Usadel equations may be solved by
matching the solutions for the two semi-infinite N--S systems;
such a solution takes the form~\cite{Altland}
\begin{equation}
   \theta_{\rm\scriptscriptstyle NS}(x,\varepsilon)
   =4\arctan\left(e^{-\kappa x} \tan {\pi\over 8}\right)\, ,
   \label{NS-solution}
\end{equation}
where $\kappa=\sqrt{-2 i \varepsilon}$ ($\re\, \kappa>0$) and the
normal layer is at $x>0$. 
Matching of such solutions was performed in 
Ref.~\onlinecite{Zharkov} by observing that for $|\theta|\ll 1$
(i.e., everywhere in the junction, except for the very thin 
layers at the interfaces) the Usadel equations become linear in
the variables $f=\theta e^{i\chi}$ and $\bar{f} =\theta e^{-i\chi}$.
Therefore the solution near the middle of the junction 
is given through the simple sum
\begin{eqnarray}
   f(x) &=& \theta_{\rm\scriptscriptstyle NS}(x)
   + e^{i\chi_0}\, \theta_{\rm\scriptscriptstyle NS}(1-x)\, , \nonumber \\
   \bar{f}(x) &=& \theta_{\rm\scriptscriptstyle NS}(x)
   + e^{-i\chi_0}\, \theta_{\rm\scriptscriptstyle NS}(1-x)\, .
\end{eqnarray}
From this solution we easily find, using (\ref{B-def}) and
(\ref{C-def}),
\begin{equation}
   C(\varepsilon)=
   32\tan^2{\pi\over 8}\, \kappa e^{-\kappa} \sin\chi_0\, ,
\end{equation}
\begin{equation}
   B(\varepsilon)= -i\varepsilon
   \left[1 - 32\tan^2{\pi\over 8}\,
   e^{-\kappa} \cos\chi_0\right]\, .
\end{equation}
With the help of the identities (\ref{integral-B}) and
(\ref{integral-C}) this immediately implies the results
(\ref{high-energy-rho}) and (\ref{C6}). The value of $C_5$ is left
undetermined by this method but may easily be obtained from
directly integrating the local density of states
(\ref{DOS-definition}) corresponding to the N--S solutions
(\ref{NS-solution}), which leads to the result (\ref{C5}).

We thank Urs Ledermann for discussions and for drawing our
attention to Ref.~\onlinecite{Zharkov} and Swiss National
Foundation for financial support.

\end{multicols}

\noindent {\bf Appendix:}

In order to verify Eq.~(\ref{integral-B}) we use the definition
(\ref{B-def}) of $B$ to re-express
\begin{eqnarray}
   &&{\partial\over\partial\varepsilon}
   (iB-2\varepsilon\hat\rho)=
   {\partial\over\partial\varepsilon}\int_0^1 dx\,
   \left[{i\over4}(\partial_x\theta)^2-\varepsilon\cos\theta+{i\over4}
   (\partial_x\chi)^2\sin^2\theta\right]
   \nonumber\\
   && = \int_0^1 dx\, \left[
   {i\over2}(\partial_x\theta)\,
   \partial_x\! \left({\partial\theta\over\partial\varepsilon}\right)
   -\cos\theta + \varepsilon\sin\theta\,
   {\partial\theta\over\partial\varepsilon}
   + {i\over2}(\partial_x\chi)\,
   \partial_x\!
   \left({\partial\chi\over\partial\varepsilon}\right) \sin^2\theta
   + {i\over2} (\partial_x\chi)^2 \sin\theta \cos\theta \,
   {\partial\theta\over\partial\varepsilon}\right]\, .
   \label{appendix-1}
\end{eqnarray}
The last integral contains five terms. After integrating the first
term by parts in $x$ (separating $\partial\theta/\partial
\varepsilon$), it vanishes against the third and fifth
terms by virtue of the first Usadel equation (\ref{Usadel-eqns}).
Integrating the fourth term in (\ref{appendix-1}) by parts in $x$
annihilates it due to the second Usadel equation; we thus arrive
at
\begin{equation}
   {\partial\over\partial\varepsilon}
   (iB-2\varepsilon\hat\rho)= -\hat\rho\, ,
\end{equation}
which is equivalent to (\ref{integral-B}). 

In order to verify
Eq.~(\ref{integral-C}) we use the identity $\partial_{\chi_0}
\int_0^1 dx\,(\partial_x\chi)^2 \sin^2\theta = 2C +\int_0^1
dx\,(\partial_x\chi)^2 \partial_{\chi_0}\sin^2\theta$
to re-express $\partial C/ \partial\varepsilon$ as
\begin{equation}
   -i{\partial C\over\partial\varepsilon}=
   -{i\over2}{\partial\over\partial\varepsilon}
   {\partial\over\partial\chi_0}\int_0^1dx\,
   (\partial_x\chi)^2\sin^2\theta +
   {i\over2}{\partial\over\partial\varepsilon}
   \int_0^1 dx\,
   (\partial_x\chi)^2 \sin 2\theta\,
   {\partial\theta\over\partial\chi_0}\, .
\end{equation}
We use the second Usadel equation to differentiate the first of
the two terms in $\varepsilon$,
\begin{equation}
   {\partial\over\partial\varepsilon}
   \int_0^1dx\, (\partial_x\chi)^2\sin^2\theta =
   2\int_0^1dx\, \partial_x\!
   \left({\partial\chi\over\partial\varepsilon}\right)
   (\partial_x\chi)\sin^2\theta
   +\int_0^1dx\, (\partial_x\chi)^2 \sin 2\theta\,
   {\partial\theta\over\partial\varepsilon}\, .
   \label{appendix-4}
\end{equation}
In the right-hand side of Eq.~(\ref{appendix-4}), the first term
is annihilated after integration by parts, and we finally arrive at
\begin{eqnarray}
   && -i{\partial C\over\partial\varepsilon}=
   -{i\over2} {\partial\over\partial\chi_0}
   \int_0^1dx\, (\partial_x\chi)^2\sin 2\theta\,
   {\partial\theta\over\partial\varepsilon}
   +{i\over2} {\partial\over\partial\varepsilon}
   \int_0^1dx\, (\partial_x\chi)^2\sin 2\theta\,
   {\partial\theta\over\partial\chi_0}
   \\
   && =
   - i {\partial\over\partial\chi_0}
   \int_0^1dx\, (\partial_x^2\theta +2i\varepsilon\sin\theta)\,
   {\partial\theta\over\partial\varepsilon}
   + i {\partial\over\partial\varepsilon}
   \int_0^1dx\, (\partial_x^2\theta +2i\varepsilon\sin\theta)\,
   {\partial\theta\over\partial\chi_0}
   = -2\int_0^1 dx\, \sin\theta\, {\partial\theta\over\partial\chi_0}
   =2{\partial\over\partial\chi_0}\hat\rho\, ,
   \nonumber
\end{eqnarray}
where we have used the first Usadel equation.

\begin{multicols}{2}

\end{multicols}


\begin{references}

\bibitem[]{}\vspace{-1cm}


\bibitem{McMillan-Belzig}
   W.~L.~McMillan,
   ``Tunneling model of the superconducting proximity effect'',
   Phys.~Rev.\ 175 (1968), 537;
   W.~Belzig, C.~Bruder, and G.~Sch\"on,
   ``Local density of states in a dirty normal metal connected to a
   superconductor'',
   Phys.~Rev.\ B {\bf 54}, 537 (1996).

\bibitem{Charlat}
   F.~Zhou, P.~Charlat, B.~Spivak, B.~Pannetier,
   ``Density of states in superconductor -- normal metal -- superconductor junctions'',
   J.~Low Temp.~Phys.\ {\bf 110} 841 (1998) [cond-mat/9707056].

\bibitem{Usadel}
   K.~D.~Usadel,
   ``Generalized diffusion equation for superconducting alloys'',
   Phys.~Rev.~Lett.\ {\bf 25}, 507 (1970).

\bibitem{Charlat-thesis}
   P.~Charlat,
   ``Transport et coh\'erence quantique dans les nanocircuits hybrides
   supraconducteur--m\'etal'',
   Ph.D.\ Thesis, Universit\'e Joseph Fourier, Grenoble (1997).

\bibitem{Bulaevskii-Buzdin}
   L.~N.~Bulaevskii, V.~V.~Kuzii, and A.~A.~Sobyanin,
   ``Superconducting system with weak coupling to the current
   in the ground state'',
   Pis'ma Zh.~Eksp.\ Teor.~Fiz.\ {\bf 25}, 314 (1977)
   [JETP Lett.\ {\bf 25}, 290 (1977)];
   A.~I.~Buzdin and M.~Yu.~Kupriyanov,
   ``Josephson junction with a ferromagnetic layer'',
   Pis'ma Zh.~Eksp.\ Teor.~Fiz.\ {\bf 53}, 308 (1991)
   [JETP Lett.\ {\bf 53}, 321 (1991)].

\bibitem{Ryazanov1}
   V.~V.~Ryazanov, V.~A.\ Oboznov, A.~Yu.\ Rusanov, A.~V.\ Veretennikov,
   A.~A.\ Golubov, J.~Aarts,
   ``Coupling of two superconductors through a ferromagnet:
   evidence for a pi-junction'',
   Phys.~Rev.~Lett.\ {\bf 86}, 2427 (2001) [cond-mat/0008364].


\bibitem{Altland}
   A.~Altland, B.~D.~Simons, and D.~Taras-Semchuk,
   ``Field theory of mesoscopic fluctuations in
   super\-cond\-uctor -- norm\-al--metal systems'',
   Adv.~Phys.\ {\bf 49}, 321 (2000) [cond-mat/9807371].

\bibitem{Ostrovsky}
   P.~M.~Ostrovsky, M.~A.~Skvortsov, and M.~V.~Feigel'man,
   ``Density of states below the Thouless gap in a mesoscopic
   SNS junction'',
   Phys.~Rev.~Lett.\ {\bf 87},  7002 (2001) [cond-mat/0012478].


\bibitem{Lukichev}
   M.~Y.~Kupriyanov and V.~F.~Lukichev,
   ``Influence of boundary transparency on the critical current of `dirty'
   SS'S structures'',
   Zh.~Eksp.\ Teor.~Fiz.\ {\bf 94}, 139 (1988)
   [Sov.\ Phys.\ JETP {\bf 67}, 1163 (1988)].

\bibitem{Zharkov}
   A.~D.~Zaikin and G.~F.~Zharkov,
   ``Theory of wide dirty SNS junctions'',
   Fiz.~Nizk.~Temp.\ {\bf 7}, 375 (1981)
   [Sov.~J.~Low Temp.~Phys.\ {\bf 7}, 184 (1981)].



\end{references}
\end{document}